\documentclass[aps,twocolumn,10pt,superscriptaddress,prl,floatfix]{revtex4-1}

\usepackage{bm}
\usepackage{graphicx}
\usepackage{subfigure}
\usepackage{natbib}
\usepackage{siunitx}

\def\be{\begin{equation}}
\def\ee{\end{equation}}
\def\e#1{\label{#1}\end{equation}}
\def\bea{\begin{eqnarray}}
\def\eea{\end{eqnarray}}
\def\ea#1{\label{#1}\end{eqnarray}}

\def\bem#1{\begin{mathletters}\label{#1}}
\def\eml{\end{mathletters}}

\def\4#1{{\boldsymbol{#1}}}
\def\8#1{{\widetilde{#1}}}
\def\bse{\begin{subequations}}
\def\ese{\end{subequations}}

\begin{document}

\title{Towards chemical structure resolution with nanoscale nuclear magnetic resonance spectroscopy}

\author{Xi Kong}
\affiliation{Institute for Quantum Optics and Center for Integrated Quantum Science and Technology, University of Ulm, D-89081 Ulm, Germany}
\affiliation{Department of Modern Physics, Hefei National Laboratory for Physics Sciences at Microscale, University of Science and Technology of China, Hefei 230026, China.}
\author{Alexander Stark}
\affiliation{Institute for Quantum Optics and Center for Integrated Quantum Science and Technology, University of Ulm, D-89081 Ulm, Germany}
\author{Jiangfeng Du}
\affiliation{Department of Modern Physics, Hefei National Laboratory for Physics Sciences at Microscale, University of Science and Technology of China, Hefei 230026, China.}
\author{Liam P. McGuinness}
\affiliation{Institute for Quantum Optics and Center for Integrated Quantum Science and Technology, University of Ulm, D-89081 Ulm, Germany}
\affiliation{School of Physics, University of Melbourne, Victoria 3010, Australia}
\author{Fedor Jelezko}
\affiliation{Institute for Quantum Optics and Center for Integrated Quantum Science and Technology, University of Ulm, D-89081 Ulm, Germany}

\begin{abstract}
Nuclear magnetic resonance (NMR) spectroscopy has approached the limit of single molecule sensitivity, however the spectral resolution is currently insufficient to obtain detailed information on chemical structure and molecular interactions. Here we demonstrate more than two orders of magnitude improvement in spectral resolution by performing correlation spectroscopy with shallow nitrogen-vacancy (NV) magnetic sensors in diamond. In principle, the resolution is sufficient to observe chemical shifts in $\sim$1\,T magnetic fields, and is currently limited by molecular diffusion at the surface. We measure oil diffusion rates of $D = 0.15 - 0.2$\,nm$^2/\mathrm{\mu}$s within (5\,nm)$^3$ volumes at the diamond surface.
\end{abstract}


\maketitle

The extension of nuclear magnetic resonance (NMR) spectroscopy to the nanoscale \cite{degen09, staudacher13, mamin13} has important implications for scientific research and technology. In particular, the attainment of sensitivities capable of detecting individual nuclei \cite{mueller14, sushkov14} has opened the way for non-destructive analysis of single molecules. In this context nitrogen-vacancy (NV) centres in diamond are particularly interesting given they have demonstrated outstanding nanoscale magnetic field sensitivity through a series of foundational experiments \cite{staudacher13, mamin13, mueller14, sushkov14, grinolds14, bala08, rondin12, zhao12, taminiau12, kolkowitz12, mcguinness13, steinert13}. However, whilst the sensitivity of NV-based NMR allows for few spin detection, the spectral resolution is below that required to observe chemical shifts, thereby severely restricting the determination of chemical structure.


The NV center is a paramagnetic defect in diamond consisting of a nitrogen impurity and a neighbouring vacancy. The spin triplet ground state possesses long coherence time and can be detected optically via coupling to optical transitions. Measuring the energy splitting between magnetic sublevels of a single NV center allows determination of the local magnetic field \cite{bala08}. The sensitivity of the measurement is then defined by the precision that the energy levels can be determined, which depends upon the coherence time and associated linewidth of the spin transition \cite{taylor08}. A narrower linewidth not only allows smaller field strengths to be detected, via shifts in the line position, but also allows finely separated energy levels to be resolved.

For NV-based sensing, closely spaced energy levels may arise from dipolar interaction between the NV electronic spin and nuclei in molecules a few nanometers away, or from the internal energy structure of such molecules. Spectral resolution below part per million levels (with respect to applied magnetic field) is vital to observing chemical shifts, quadrupole interactions and dipolar couplings which allow molecular structure to be derived. Likewise, the identification of hyperfine interactions induced by NV centers at few nanometers distance from nuclear spins requires sub-kilohertz spectroscopy. Therefore prolongation of the NV sensing time beyond this limit is important to enable applications of nanoscale sensing to structural analysis, and is especially relevant for centers located close to the diamond surface.

\begin{figure*}[tbh]
\begin{center}
\includegraphics[width=0.9 \textwidth]{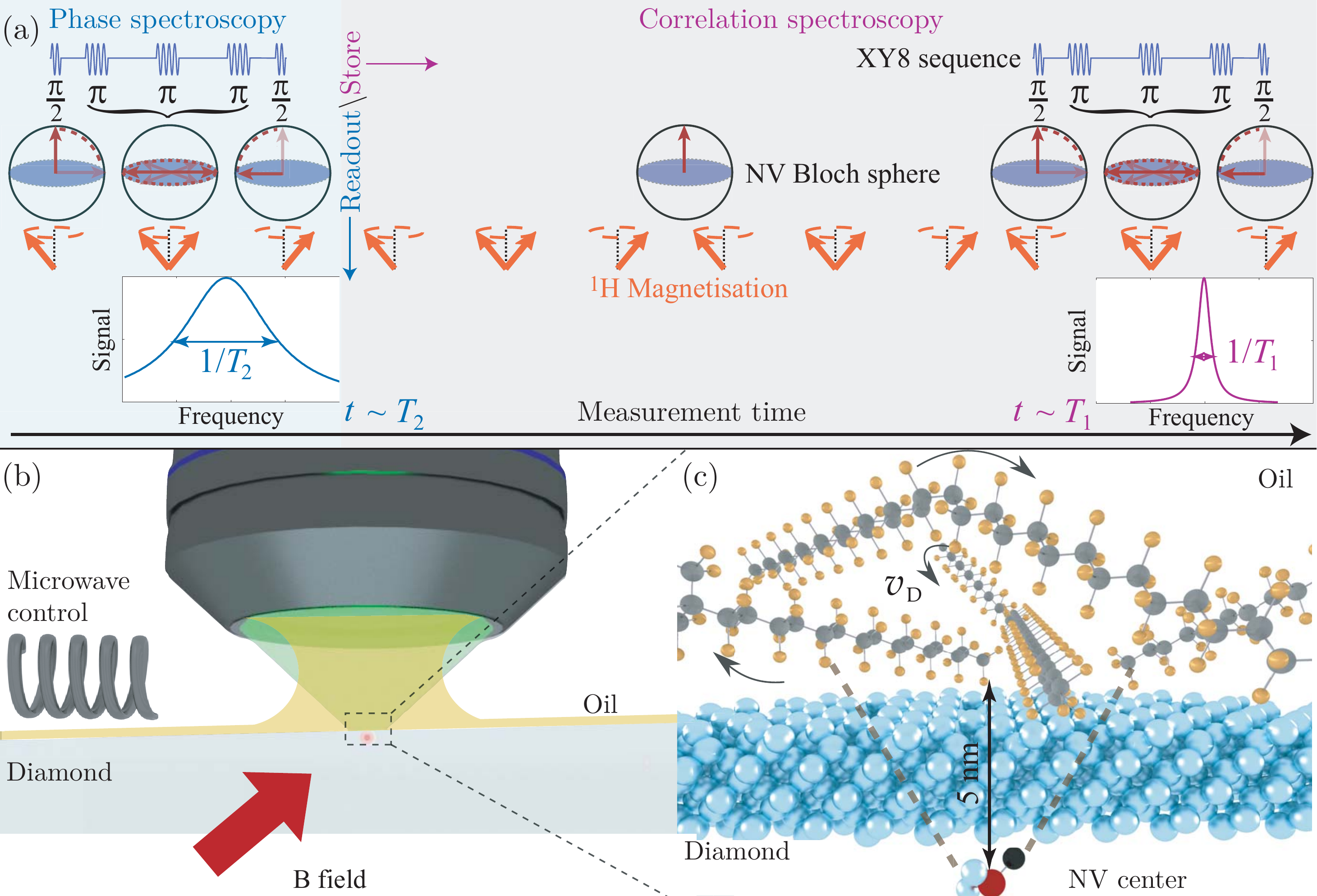}
\protect\caption{Nanoscale magnetic sensing based on coherent phase and spin population timescales. (a) Magnetometry protocol using coherent phase accumulation: the sensing qubit is driven/pulsed at a desired frequency with e.g. an XY8 sequence. Evolution around the Bloch sphere is altered by magnetic fields at the sensing frequency, leading to differences in phase accumulation and finally spin population upon readout. Spectral resolution depends on the $T_2$ time of the sensor. Protocol using spin population correlations: an initial phase accumulation is stored in the sensing qubit’s spin population. A second phase accumulation is correlated with the initial measurement by a second mapping to spin population. The final population is a coherent sum of two measurements, dependent on the phase evolution of the external field during time, $t$. (b) A single nitrogen vacancy spin qubit below the diamond surface is used to measure the NMR spectrum of protons in a nanoscale volume of oil molecules. A magnetic field is applied with a permanent magnet and microwaves are carried by a copper wire. (c) Due to the extremely near-field interaction between the sensing qubit and sample, the recorded NMR spectrum is highly dependent on the NV distance to the surface. Diffusional motion and interactions near the surface of the diamond affect the recorded spectrum.} \label{fig:intro}
	\end{center}
\end{figure*}

One approach to improve NV magnetic resonance spectroscopy is to remove decoherence sources which act to limit the detection time, either using quantum control techniques (e.g. dynamical decoupling) or by physical treatment (e.g. sample cooling and purification). Recently nanoscale NMR with shallow NV centers was realized, with spectral linewidths limited by the NV dephasing ($T_2$) time \cite{staudacher13, mamin13, mueller14}. Dynamical decoupling protocols were implemented to extend the NV dephasing time, resulting in a commensurate increase in sensitivity and spectral resolution. Dynamic decoupling alters the phase evolution of the sensing qubit such that the effect of unwanted noise is removed, whilst retaining sensitivity to signals at a particular frequency. Pulsed decoupling techniques (e.g. XY-N, CPMG-N) operate by filtering signals at harmonics of the pulsing period whereas continuous decoupling (e.g. spin-locking, Hartmann-Hahn polarization transfer) tune the driving frequency of the qubit [Fig.\,\ref{fig:intro}(a)]. Limitations in driving strength and fidelity however, reduce the effectiveness of decoupling, and provide imperfect filtering. For example spectral noise faster than the decoupling speed is not removed, leading to proposals based on quantum error correction \cite{arrad14, kessler14}. For NV sensing this problem is particularly salient, as poor scaling of dephasing time with number of applied pulses has been observed for shallow NV centers \cite{romach15, myers14}. As a result spectral resolution on the order of 10 -- 100 kHz remains the benchmark, which is more than an order of magnitude worse than the intrinsic NV relaxation ($T_1$) time.

An alternative avenue towards high resolution spectroscopy is to transfer the signal encoded in the NV phase to spin population and then rely upon the long $T_1$ time of the NV center. The idea of extending resolution to timescales limited solely by spin lattice relaxation is closely related to stimulated echo techniques \cite{mims72} and was recently demonstrated for coherently coupled spins in diamond \cite{laraoui11, laraoui13}. Although this approach does not enhance the sensitivity of the NV center which relies upon coherent phase accumulation, spectral resolution can, in principle, be extended to the $T_1$ limit [Fig.\,\ref{fig:intro}(a)]. During the first ``preparation'' sequence consisting of a multipulse (e.g. XY8) echo, the magnetisation of precessing nuclear spins is mapped onto the NV electronic spin population. After an evolution time $t$, during which the phase of the nuclear magnetisation progresses, the subsequent nuclear phase is then correlated with the initial phase by a second ``read'' echo sequence which again maps to the NV spin population \cite{SuppMat}. The resultant signal is a modulation at the frequency of nuclear precession.

\begin{figure}
\includegraphics[width=0.9 \columnwidth]{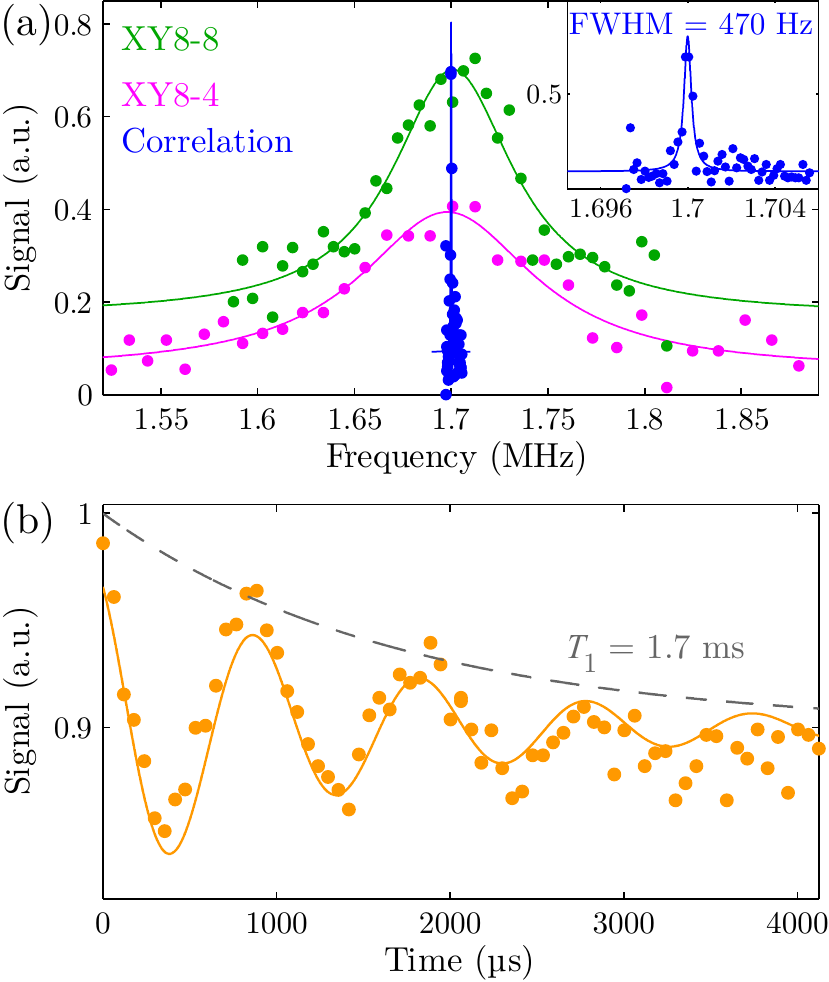}
\protect\caption{Comparison between coherent phase spectroscopy and correlation spectroscopy. (a) Proton NMR spectra obtained with XY8-4 (magenta), and XY8-8 (green) dynamical decoupling, giving a spectral linewidth of 110 kHz and 74 kHz respectively. The spectrum of an externally applied AC field, measured with correlation spectroscopy (blue) has a corresponding linewidth of 0.47 kHz (inset). (b) Plot of data obtained with correlation spectroscopy before Fourier transform. The 1.7 MHz AC field was under-sampled in order to observe the envelope of the correlation signal. The $T_1$ decay of the NV center ($T_1 = 1.7$\,ms) is shown as the grey dotted-line.} \label{fig:corr}
\end{figure}

A schematic of the measurement is shown in Fig.\,\ref{fig:intro}(b). We use shallow NV centers implanted 2--5\,nm into a $\langle 100 \rangle$ diamond surface by 2.5\,keV N$^+$ implantation. The diamond is 99.999\% $^{12}$C isotopically enriched, with $\sim$10 part per billion impurity content so that the NV spectral environment is dominated by magnetic species at the surface or outside of the diamond \cite{romach15}. A 400 Gauss magnetic field was applied along the NV axis ($\langle 111 \rangle$ diamond crystal axis), resulting in a proton Larmor frequency of 1.7\,MHz ($^1$H gyromagnetic ratio, $\gamma_{\mathrm{H}} = 4.25$\,kHz/Gauss, note the use of real frequency, not angular frequency values). The NMR spectrum of statistically polarized protons in an oil layer placed on the diamond surface was then measured with an XY8 sequence \cite{staudacher13}. A peak in the spectrum corresponding with the $^1$H Larmor frequency records the root-mean squared magnetic field amplitude of the protons, as seen by the NV center [Fig.\,\ref{fig:corr}(a)]. From the amplitude of the nuclear spin signal \cite{SuppMat}, we calculate the NV distance to the diamond surface as $d = 5$\,nm, with a detection volume of (5\,nm)$^3$ [Fig.\,\ref{fig:intro}(c)].

The resolution of the spectra shown in Fig.\,\ref{fig:corr}(a) is initially determined by the NV dephasing time. For an XY8-4 sequence consisting of 32 decoupling pulses, the measurement filter function has a linewidth of 110\,kHz due to the $\sim 10$\,$\mu$s NV $T_2$ time. By increasing the number of decoupling pulses the NV $T_2$ time could be increased, leading to a narrower NMR signal with 74\,kHz linewidth [Fig.\,\ref{fig:corr}(a)]. However the resolution of these initial measurements remains poorer than the expected proton linewidth (typically in the tens of kilohertz range for solid-state NMR and less for liquid-state NMR), meaning little information about chemical structure or sample relaxation rates can be obtained. We therefore performed correlation spectroscopy out to millisecond timescales, allowing a significant improvement in spectral resolution.

As a demonstration of the technique we measured an effective delta frequency signal from a high-stability signal generator. An XY8 sequence followed by a variable time delay and a second XY8 sequence resulted in oscillations of the NV population at the signal generator frequency \cite{SuppMat}. In order to detect the correlation oscillations over long evolution times, whilst limiting the number of datapoints required, we applied an under-sampling protocol which recorded the envelope of the correlation signal. The sampling rate was chosen to be 17\,kHz (i.e. a sample every 58.859\,$\mu$s) between $2f_H/n$ and $2f_L/(n-1)$, with  $f_H=1.704$\,MHz, $f_L=1.699$\,MHz and $n=201$ to fulfil the Nyquist criterion \cite{oppenheim75}. The result is shown in Fig.\,\ref{fig:corr}(b), from which we obtained the linewidth plotted in Fig.\,\ref{fig:corr}(a) by Fourier transform. A resolution improvement of over two orders of magnitude to 470\,$\pm$\,40\,Hz can be seen, in comparison to coherent phase sensing measurements.

To our knowledge, the spectral resolution demonstrated here is a record for magnetometry with shallow NV centers \cite{staudacher13, mueller14, romach15, loretz14, rosskopf14, myers14}. Not only does the resolution outperform current nanoscale NMR techniques, but it is comparable to measurements performed on protected NV centers deep inside diamond \cite{laraoui13, maurer12, bala09}. To examine whether the linewidth is indeed limited by spin lifetime, we independently measured the $T_1$ time of the NV center. We find the $T_1$ decay of 1.7\,ms closely matches the damping of the correlation signal [Fig.\,\ref{fig:corr}(b)]. The good agreement between $T_1$ decay and the measured linewidth indicates that it is the timescale which information is stored in the NV spin population rather than drifts in the external magnetic field or experimental setup, that limit the resolution of these measurements.

\begin{figure}
\includegraphics[width=0.9 \columnwidth]{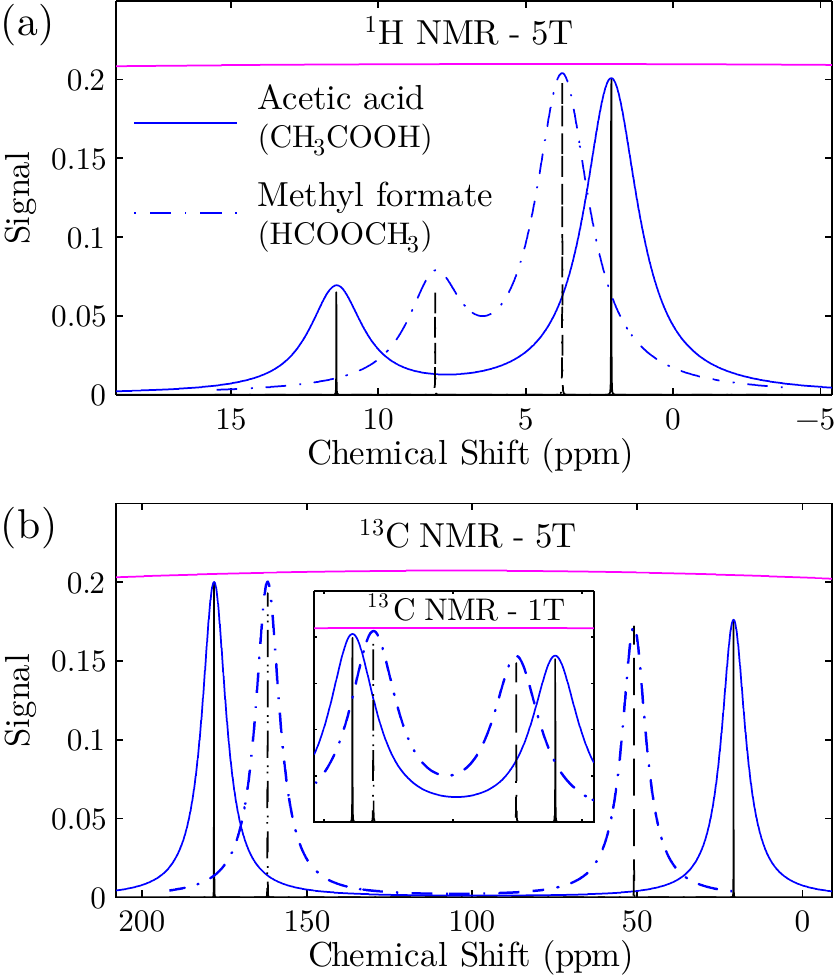}
\protect\caption{
Chemical shift resolution with an NV sensor. (a) Simulated $^1$H NMR spectra of acetic acid and methyl formate at 5\,T with 470 Hz resolution from correlation spectroscopy (blue) and 110 kHz resolution from phase spectroscopy (magenta). (b) Simulated $^{13}$C NMR spectra at 5\,T with correlation and phase spectroscopy. (Inset) Identical NMR spectra as (b), but for 1\,T magnetic field, resulting in factor of five poorer resolution. Note, the spectra only display the intrinsic resolution of the NV sensor, sample relaxation/diffusion is not taken into account.}
\label{fig:sims}
\end{figure}

NMR spectroscopy with few hundred hertz resolution is capable of identifying molecular structure in moderate magnetic fields based on chemical shifts. In Fig.\,\ref{fig:sims} we show the simulated $^1$H and $^{13}$C NMR spectra for acetic acid and methyl formate with 470\,Hz resolution, where we have neglected any intrinsic sample relaxation and diffusion. The two molecules have identical chemical makeup, differing only in structure, meaning they cannot be distinguished by techniques sensitive to proton/carbon ratio. We show in 5 Tesla magnetic fields, the chemical shift arising from different functional groups can be resolved, which allows for unambiguous identification by comparison to NMR databases \cite{NMR_database}. By comparison, coherent phase spectroscopy is unable to yield any structural resolution due to the poorer linewidth [Fig.\,\ref{fig:sims} (magenta)]. The higher chemical shift experienced by carbon atoms mean that chemical shifts can be resolved in magnetic fields of 1 Tesla, making $^{13}$C NMR a more attractive option for NV based spectroscopy [Fig.\,\ref{fig:sims} (b, inset)].

Acetic acid and methyl formate were chosen as they contain carboxylic acid and ester functional groups which are present at the diamond surface. Such groups are targets for diamond functionalisation and expected to play a role in the charge state and decoherence of shallow NV centers. We note that the realisation of chemical shift spectroscopy is predicated on the ability to both readout the NV spin state and preserve long $T_1$ times at high magnetic fields, both of which have been demonstrated recently \cite{stepanov15, aslam15}. Additionally, the intrinsic sample relaxation rates must be low enough to allow for high resolution spectroscopy.

\begin{figure}
\includegraphics[width=0.9 \columnwidth]{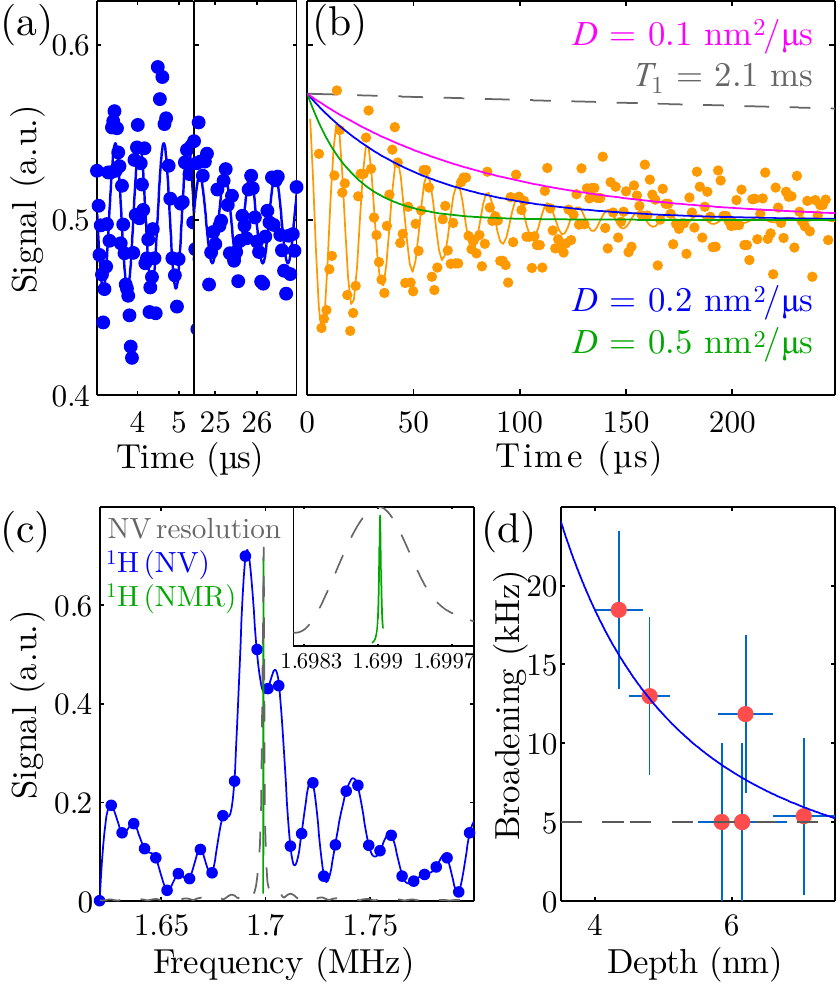}
\protect\caption{(a) Nanoscale proton correlation spectroscopy. (a) Oscillations due to Larmor precession of protons near the NV sensor, shown at two time intervals. (b) Long-time correlation signal obtained by under-sampling. The $T_1$ decay of the NV center ($T_1 = 2.1$\,ms) is shown as the grey dotted-line. Simulations comparing the decay rate for an oil diffusivity of 0.1 (magenta), 0.2 (blue), and 0.5\,nm$^2/\mathrm{\mu}$s (green) with an oil density of 1.09\,g/mL and a proton nuclear spin density of 50\,nm$^{-3}$. (c) Fourier transform of the correlation signal, showing the measured proton NMR linewidth of 20\,kHz (blue) compared to the intrinsic linewidth of the NV sensor (grey dotted-line), and conventional NMR proton spectrum of oil (green). (d) Linewidth broadening vs depth for 6 NV centers (red datapoints), and fit to data with $D$ = 0.15 nm$^2/\mathrm{\mu}$s (blue line), measured with XY8 spectroscopy. Minimum detectable broadening (5\,kHz) shown as dotted line.
}
\label{fig:proton}
\end{figure}

In order to determine whether detection of chemical shifts is indeed readily achievable, and to investigate relaxation mechanisms at the diamond surface we performed correlation spectroscopy of protons at the diamond surface with another NV center of similar depth. Fig.\,\ref{fig:proton} (a) shows the signal measured to 27\,$\mu$s, with characteristic oscillations at the $^1$H Larmor frequency. Again, we applied an under-sampling protocol to record the envelope of the correlation signal. The sampling rate was chosen to be 0.81\,MHz between $2f_H/n$ and $2f_L/(n-1)$, with  $f_H=1.75$\,MHz, $f_L=1.65$\,MHz and $n=5$. Oscillations continue beyond 100\,$\mu$s, which although longer than could be investigated with XY8 decoupling, is much shorter than the $T_1$ time of this NV center [Fig.\,\ref{fig:proton} (b)]. From the Fourier transform of the signal we determine the proton linewidth to be 20\,kHz, which is now no longer limited by the NV dephasing time [Fig.\,\ref{fig:proton} (c)].

Interestingly the NV measured proton spectrum is significantly broader than the $\sim 40$\,Hz width spectrum measured by conventional NMR [Fig.\,\ref{fig:proton} (c)] (see \cite{SuppMat} for details). In nanoscale NMR, molecules can leave the detection volume by molecular diffusion which reduces interaction time with the sensor and results in a broadened linewidth. In contrast, molecular diffusion in conventional NMR leads to narrower linewidths through motional narrowing. To test whether diffusion is responsible for the linewidth broadening, we measured the proton linewidth with six NV centers at different depths using XY8 spectroscopy and recorded broadening (beyond the measurement resolution) as a function of sensor depth [Fig.\,\ref{fig:proton} (d)]. Despite the lower resolution associated with using XY8 spectroscopy, we observe a clear increase in spectral linewidth as the NV depth reduces, which is fitted by a pure diffusion model ($\Delta \omega = 2D/d^2$, where $\Delta \omega$ is broadening) to give a diffusion constant, $D = 0.15\pm0.04$\,nm$^2/\mathrm{\mu}$s. We note that our model does not take into account nuclear-nuclear interactions, interactions with spins at the diamond surface, or a static boundary layer of non-diffusing liquid which would be expected to give a different depth scaling.

To validate the diffusional model we numerically simulated the magnetic field produced by diffusing protons at the diamond surface. We used a cluster correlation expansion to determine the magnetic field produced by 3000 randomly moving nuclear spin pairs, with their motion calculated in 0.1\,$\mu$s timesteps \cite{SuppMat}. As shown in Fig.\,\ref{fig:proton} (b), a diffusion coefficient of $0.19\pm0.05$\,nm$^2/\mu$s, gives a best fit to the experimental results. Calculation based on pure diffusion, $\overline{x^2} \simeq 2D t$, where $\overline{x^2}$ is the mean squared displacement, reproduces this value well for diffusion through a (5\,nm)$^3$ sensing volume, with a decay timescale of 65\,$\mu$s, in agreement with Fig.\,\ref{fig:proton} (b).


In summary, we have demonstrated that correlation spectroscopy can be applied to nanoscale NMR for improving the spectral resolution of diamond magnetometry beyond values limited by the coherence time of electron spins. The protocol is limited by the longitudinal relaxation time of the NV spin (several milliseconds at room temperature) leading to sub-kilohertz resolution, and in principle, allowing chemical shifts to be observed in high magnetic field experiments. Importantly, by storing information in spin population, susceptibility to low frequency noise is reduced, which has been shown to be dominant at the diamond surface \cite{romach15, myers14, rosskopf14}. In contrast to related relaxation spectroscopy techniques \cite{steinert13, kaufmann13}, the linewidth does not depend on the NV $T_2^*$ time.

We have also shown how molecular motion at the diamond surface affects the recorded spectrum, which can be used to determine molecular diffusion rates. In addition we highlight the importance of chemical attachment in order to achieve high resolution spectra. Improvements in stabilisation of molecules to the diamond surface, coupled with correlation spectroscopy at high fields, promise to yield information on molecular structure, morphology and dynamics at the single molecule level.


\bibliography{NV}

\end{document}